# A hierarchical approach for pulmonary nodules identification from CT images using YOLO v5s nodule detection and 3D neural network classifier

Yashar Ahmadyar Razlighi[1], Alireza Kamali-Asl[1], Hossein Arabi[2]

[1]Department of Medical Radiation Engineering, Shahid Beheshti University, Tehran, Iran

[2]Division of Nuclear Medicine & Molecular Imaging, Geneva University Hospital, CH-1211, Geneva, Switzerland

**Abstract**

**Purpose:** Lung cancer is one of the leading causes of death worldwide. Early lung cancer detection would enable physicians to treat patients effectively and decrease mortality rates significantly. This work proposes a hierarchical framework for lung nodule detection and classification from computed tomography (CT) images.

**Method:** In the first step, a pre-trained model (YOLO) was used to detect all suspicious nodules. The YOLO model was re-trained using 397 CT images to detect the entire nodule in CT images. To maximize the sensitivity of the model, a confidence level (the probability threshold for object detection) of 0.3 was set for nodule detection in the first phase (ensuring the entire suspicious nodules are detected from the input CT images). The aim of the hierarchy model is to detect and classify the entire lung nodules (from CT images) with a low false-negative rate. Given the outcome of the first step, we proposed a 3D CNN classifier to analyze and classify the suspicious nodules detected by the YOLO model to achieve a nodule detection framework with a very low false-negative rate. This framework was evaluated using the LUNA 16 dataset, which consists of 888 CT images containing the location of 1186 nodules and 400000 non-nodules in the lung.

**Results:**

A large number of false positives were detected due to the low confidence level used in the YOLO model. Utilizing the 3D classifier, the accuracy of nodule detection was remarkably enhanced. The YOLO model detected 294 suspicious nodules (out of 321) when using a confidence level of 50%, wherein there were 107 false positives (187 true positives). By reducing the confidence level to 30%, 459 suspicious nodules were identified by the YOLO model, wherein 138 were false positives, and 321 were true positives. When the outcome of the YOLO model with a confidence level of 30% was fed into the 3D CNN classifier, a nodule detection accuracy of 98.4% and AUC of 98.9% were achieved

**Conclusion:** The proposed framework resulted in a few false-negative and false-positives predictions in nodule detection from CT images. The proposed approach would be helpful in detecting pulmonary nodules from CT images as a decision support tool.

**Keywords:** Lung cancer, pulmonary nodules, object detection, classification, CT, deep learning.

# INTRODUCTION

Lung cancer is one of the leading causes of death in the world. The lung cancer research foundation predicts that in 2022, there will be 236,740 new cases of lung cancer and 130,180 deaths from lung cancer in the United State. Approximately 540,000 Americans have been diagnosed with lung cancer at some point in their lives [1]. The detection of pulmonary nodules is an essential step in detecting lung cancer at an early stage. A lung nodule is an abnormal growth that occurs in the lung, wherein small nodules (less than 9 mm) are usually not cancerous, but they may still be signs of an early stage of cancer [2]. Early lung cancer detection is feasible through identifying these nodules.

Lung cancer screening can be conducted in a variety of ways, including bronchoscopy, which is a procedure that allows physicians to explore the inside of the airways and take samples of cells. These biopsy samples are analyzed in a laboratory to detect abnormal cells. Another way is CT scan-guided biopsy, wherein for nodules on the outer portion of the lung, CT images are employed to guide a thin needle through the skin and into the lung. The purpose of this procedure is to obtain tissue samples from the nodule in order to examine them for abnormalities. Moreover, Positron Emission Tomography (PET) imaging is also exploited to detect cancerous cells in organs [3].

United State preventive services recommend that individuals at risk for lung cancer undergo a low-dose CT scan every year. A low-dose chest computed tomography scan (LDCT) may be able to improve the overall survival rate by detecting lung cancer in its early stages. There has been a renewed interest in lung cancer screening due to the widespread use of LDCT imaging [4-7].

Artificial intelligence algorithms, including deep convolutional neural networks (DCNNs), have been widely used to detect lung nodules automatically [8-10]. In recent years, researchers have increasingly used 3D Convolutional Neural Networks (CNNs) for nodule detection from CT images using a segmentation framework by the UNet networks, detecting/delineating the nodules in the lung [11-16]. Thereafter, to enhance the overall accuracy, a second classifier would analyze the segmented nodules to detect false-positive and -negative cases [17-20]. For a faster and more accurate approach, Mask-R-CNN has been used, which is a two-stage object detector (Region Proposal Network (RPN)) followed by a region-based convolutional neural network (R-CNN) and a semantic segmentation model (MASK) [21-23]. In the first step, a bounding box encompassing the target object is defined through the Selective Search approach. In the next step, a CNN layer is applied to classify the detected objects.

Nguyen et al. used Fast R-CNN model with an adaptive anchor box for lung nodule detection in CT images. The Fast R-CNN model was trained using nodules with different sizes in the training dataset and adaptive anchor boxes with different sizes. Normally R-CNN is used with a fixed anchor box, but here the size of the anchor box varied to match the nodules' size for better detection accuracy. Concerning false positives reduction from the Fast R-CNN's output, a residual convolutional neural network architecture (ResNet) [24-27] was proposed to post-process the detected nodules. This method was trained and tested on the LUNA16 dataset [28, 29].

In another study, a hierarchical method consisting of R-CNN was proposed for nodule detection, and a 3D ResNet was employed for false positive reduction [30]. Due to the slow processing of the R-CNN models, these approaches are not suitable for real-time applications [31]. Furthermore, Agnes et al. [32] employed a two-stage model consisting of a UNet-based network (Atrous Unet+) for node detection and a pyramid-dilated convolutional LSTM network for false positive reduction.

YOLO (You Only Look Once) [33] is one of the most prominent object detection algorithms and is able to perform real-time object detection using a single forward pass. It detects the object and classifies the detected objects according to the labels simultaneously. One of the first medical applications of YOLO was to detect and classify breast masses in mammography images [34-37]. Additionally, it has been used to detect skin cancer [35] and melanoma [38]. In this regard, by combining DetectNet and GooleNet architectures, George et al. [39] presented an object detection method for detecting pulmonary nodules in CT images. The DetectNet, used in the abovementioned work, is based on the YOLO architecture [40], which explores the whole image for suspicious lesion detection and classification into nodule or non-nodule cases. In another approach proposed by Huang et al. [41] lung nodule detection was conducted using a 3-D OSAF-YOLOv3 model. This model is an integration of the 3-D YOLOv3 with the one-shot aggregation module (OSA), the receptive field block (RFB), and a feature fusion scheme (FFS). Despite the promising performance of the YOLO-based nodule detection model, it may not always be accurate, and a significant number of nodules may remain undetected (high false positives). This shortcoming could be addressed through introducing a compartment to the output of the model to reduce the number of false positives.

In this study, a hierarchical approach is proposed to detect and classify nodules in the lung. Initially, suspicious nodules will be detected using an object detection framework. Then these nodules will be further processed to determine whether they are nodules or non-nodules (using a 3D convolutional classifier). In the first step, a pre-trained YOLO model re-trained for nodule detection on CT images, was used to detect the entire suspicious nodules. The goal of this step is to detect the entire suspicious nodules. To minimize the false-negative and false-positive rates in the detected nodule by the YOLO model, a 3D CNN classifier was proposed and incorporated into the framework to classify the suspicious nodules accurately.

## 2. MATERIAL AND METHODS

### 2.1 Overview

This model, referred to as hierarchical nodule detection (HND), has two stages. In the first stage, the entire CT image is analyzed using the YOLO algorithm, which is a commonly used algorithm for object detection. The YOLO algorithm would determine the location of the nodules and the probability (confidence score) of being nodules. Given the locations of the entire suspicious nodules detected by the

YOLO network (using a low confidence score), a 3D bounding box (containing the nodule and background tissue) can be defined around each nodule to be fed to the next module. In the next stage, a 3D CNN classifier will process these 3D patches to determine whether or not the candidates are real nodules. In short, the first stage with YOLO algorithm would detect the entire nodule as a course classification, and the second module will perform a fine classification by focusing on a single nodule at a time (Figure 1). The aim of the 3D CNN classifier is to minimize the false negative and false positive rates.

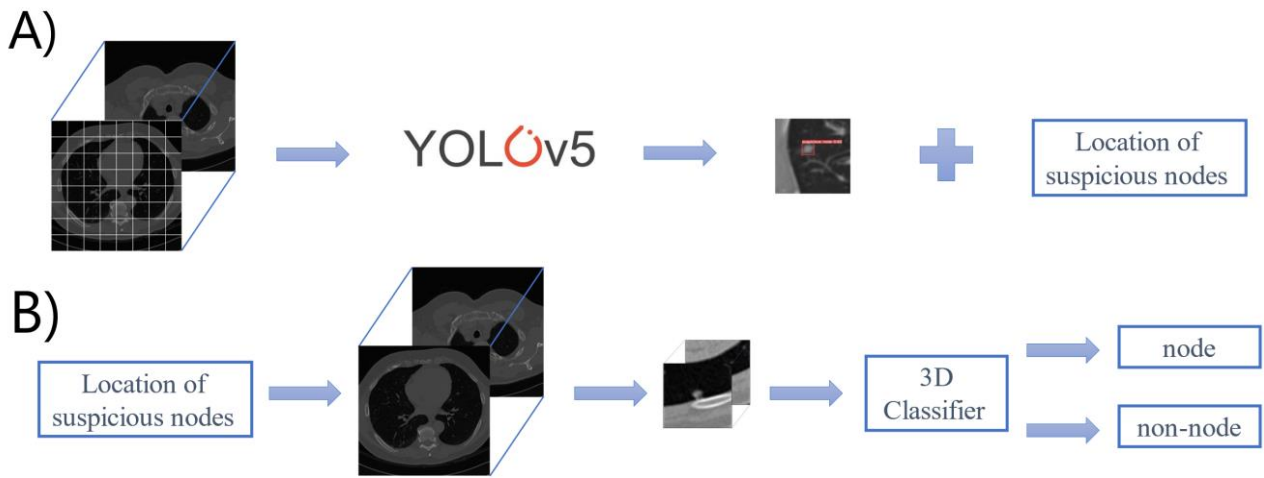

**Figure 1**. The proposed framework (HND) for the detection of cancerous nodules in CT images. The entire 3D CT image is fed into the YOLO model in order to determine the location of the entire suspicious nodules (A). Given the location of the suspicious nodules, 3D patches of the image are fed into the 3D classifier to classify them as nodules or non-nodule (B).

## 2.2 Dataset and Preprocessing

In order to assess the performance of the proposed deep learning-based framework, 888 CT images together with labeled nodules from the Lung Nodule Analysis Challenge 2016 (LUNA 16) [29] dataset were employed. The dataset represents a subset of the LIDC-IDRI dataset, in which each subject includes a CT image of 512×512×100-500 voxel. A team of four expert radiologists examined the CT images and classified the nodules and non-nodules. At least three out of four radiologists must have labeled positive to be considered as a nodule (having a diameter of at least 3 mm). In total, there are 1186 nodules in this dataset. 597 patients have been chosen which consist of most of the nodules (1125). The entire CT images were converted into HU and resampled into an isotropic voxel size of 1 mm. The CT images were normalized to an intensity range of 0 to 1 using a global normalization factor (maximum intensity in the entire dataset).

**2.3 Network training**

*1) YOLO network*

The YOLO network was re-trained to find the nodules within the whole input CT image. YOLO training was performed in a 2D mode to explore the entire slices. Regarding the small sizes of nodules compared to 2D slices of the CT images, CT slices were divided into 64×64 sub-images (Figure 2). The sub-images containing nodule tissue were employed to train the YOLO model.

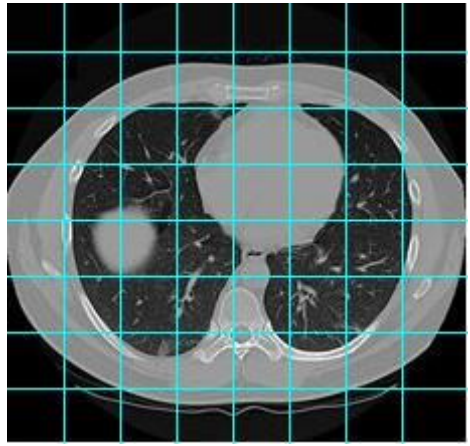

**Figure 2**. A transaxial slice of CT image, divided into 64 ×64 sub-images for the training of the YOLO network (The sub-images containing lung tissue were employed to train the YOLO model).

A pre-trained YOLO model on the COCO dataset, which is an object detection, segmentation, and captioning dataset [42], was re-trained for nodule detection. YOLO is based on a CNN architecture [43], which consists of input, backbone, neck, and output (Figure 3). The input terminal performs the primary preprocessing, while the cross-stage partial networks (CSP) [44] and Spatial Pyramid Pooling (SPP), the backbone of the network, are used to extract features from input data. The neck is constructed using the feature pyramid network (FPN) and the pixel aggregation network (PAN). The FPN transfers comprehensive semantic information from the top to the lower feature map. The feature maps from lower to higher level localization are conveyed by PAN [45]. When these two structures are combined, they enhance the features extracted by the backbone and improve the overall performance of object detection. As a final step, the output layers are used to detect objects with different sizes based on the feature maps. A re-training was done for the YOLO algorithm to identify all nodules in the CT images regardless of whether they are nodules or non-nodule.

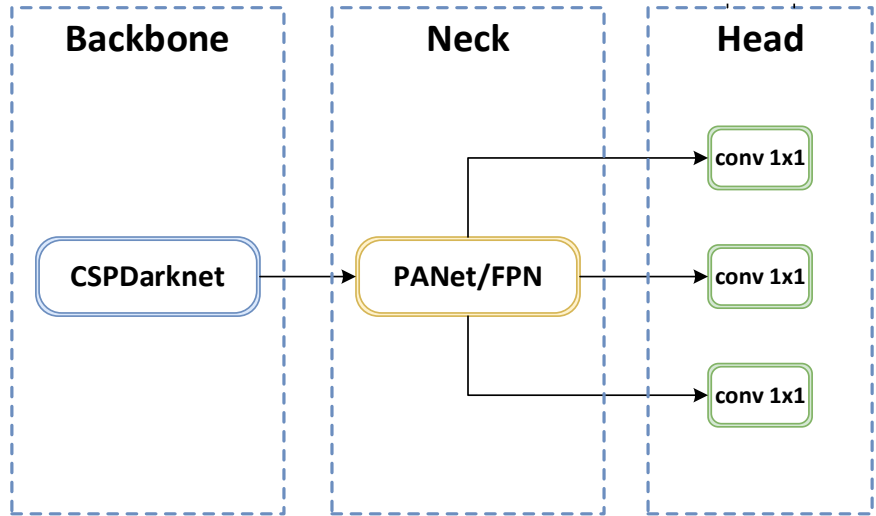

**Figure 3**. Architecture of YOLO network.

The re-training of the YOLO model was performed utilizing 804 nodules in 397 patients in 300 epochs with SGD (stochastic gradient descent), an initial learning rate of 0.01, and a batch size of 16. YOLO output is a bounding box around the detected object, a confidence score, and the center of the nodules. The confidence score indicates the probability of object detection formulated as $C = Pr(object) \times IoU$, wherein *IoU* stands for intersection over the union between the predicted box and ground truth. In order to ensure the detection of entire nodules in the input CT image, the confidence level of the YOLO model was reduced from 0.5 to 0.3.

*2) 3D CNN network*

Regarding the small size of the lung nodules in the lung, a patch of 64×64×64 voxels was defined around each nodule for the training of the 3D CNN network. These patches are fed into the 3D CNN network to be classified as nodule and non-nodule. In the proposed 3D CNN classifier, there are four convolutional layers, where the number of filters increases from 64 in the first layer to 256 in the last. In addition, 3D Maxpooling and batch normalization layers were applied after each convolutional layer. In order to prevent overfitting, two dropout layers have been considered before the output. The ReLu function was used as an activation function for the internal layers, and a sigmoid function was used together with the dense layer for the binary classification. (Figure 4).

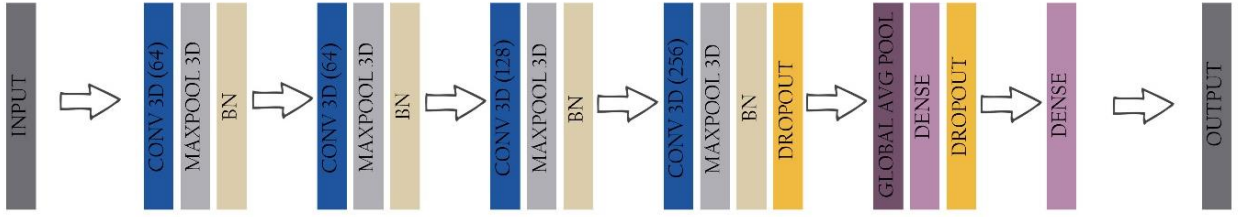

**Figure 4.** Architecture of the proposed 3D CNN for nodule classification.

For training of the 3D CNN model, a learning rate of 0.0001, together with a decay factor of 0.96 was employed. The optimization process was achieved using the Adam optimizer, and we used binary cross entropy as a loss function. The model was trained in 100 epochs with a batch size of 64. Same as YOLO, 804 nodules were employed for the training of this network. This network was trained and evaluated using the outputs of the YOLO model.

**2.3 Evaluation strategy**

In order to evaluate the model, we first identified the suspicious nodules using a confidence level of 0.5 for the YOLO model. Then, to ensure the detection of the entire suspicious nodules, a confidence level of 0.3 was examined. Given the suspicious nodules detected by the YOLO model, the patches containing the suspicious nodules were fed into the 3D classifier. This process was done for 200 CT images (200 patients) containing 321 nodules as a test dataset for the evaluation of the model.

The performance of the model in nodule classification was assessed using the accuracy score, precision score, recall score, and f1-score as follows:

*Accuracy Score = (TP + TN)/ (TP + FN + TN + FP)* (1)

*Precision Score = TP / (FP + TP)* (2)

*Recall (sensitivity) Score = TP / (FN + TP)* (3)

*F1 Score = 2× Precision Score × Recall Score/ (Precision Score + Recall Score)* (4)

*(True Positive = TP, False Positive = FP, True Negative = TN, False Negative = FN)* (5)

Moreover, receiver operating characteristics (ROC) was plotted using the true positive rate (TPR) versus the false positive rate (FPR) at different thresholds, and the area under this curve (AUC) was calculated.

*TPR = TP / (TP+FN)* (6)

*FPR = FP / (FP+TN)* (7)

## 3. RESULTS

The YOLO model was evaluated on 200 CT images containing 321 nodules. The YOLO model identified 294 objects out of 321 real nodules as suspicious nodules, using a confidence score of 0.5 (107 were false positives). Then it detected 459 suspicious nodules out of 321 real nodules using a confidence score of 0.3, wherein 138 were false positives (FP) (28% FP rate). This could be justified by the low confidence score used in the YOLO model, and the fact that YOLO is an effective algorithm for large objects not tiny structures like nodules. The outputs of the evaluation processes have been shown in Table 1. Figure 5 depicts a representative true positive and false positive detected by the YOLO model. To reduce the high false positive rate, the output of the YOLO model was fed into the 3D CNN classifier, and the precision of the nodule detection was improved from 69% to 100% by the 3D CNN classifier discriminating between nodules and non-nodules. The ROC plot for the model is presented in Figure 6. In addition, Figure 7, Table 2, and Table 3 present the outcomes of the 3D CNN model on YOLO outputs with a 0.3 confidence level (HND model).

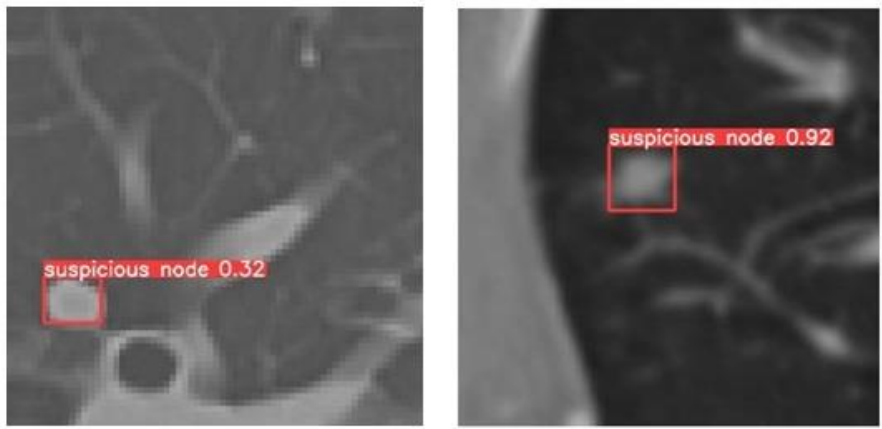

**Figure 5.** Representative false positive (left) and true positive (right) nodules detected by the YOLO model.

Using a confidence level of 0.3, the entire nodules (321 real nodules) were detected in the CT images of 200 patients.

**TABLE 1.** Outcome of the YOLO model before being processed by the 3D CNN model.

| Confidence Level | Number of real nodules | True Positive | False Positive | Precision(%) | Detected nodule per real nodule (%) |
|---|---|---|---|---|---|
| **0.5** | 321 | 187 | 107 | 63% | 58% |
| **0.3** | 321 | 321 | 138 | 69% | 100% |

**TABLE 2.** Results of the 3D classifier on the outputs of YOLO with a confidence level of 0.3.

| Accuracy(%) | Recall(%) | Precision(%) | AUC(%) |
|---|---|---|---|
| 98.4 | 97.8 | 100 | 98.9 |

**TABLE 3.** Results of the HND (YOLO 0.3+ 3D Classifier) model evaluated for nodules and non-nodules.

| Class | Precision(%) | Recall(%) | f1-score(%) |
|---|---|---|---|
| **Nodule** | 100 | 97.8 | 99 |
| **Non-Nodule** | 95 | 100 | 98 |

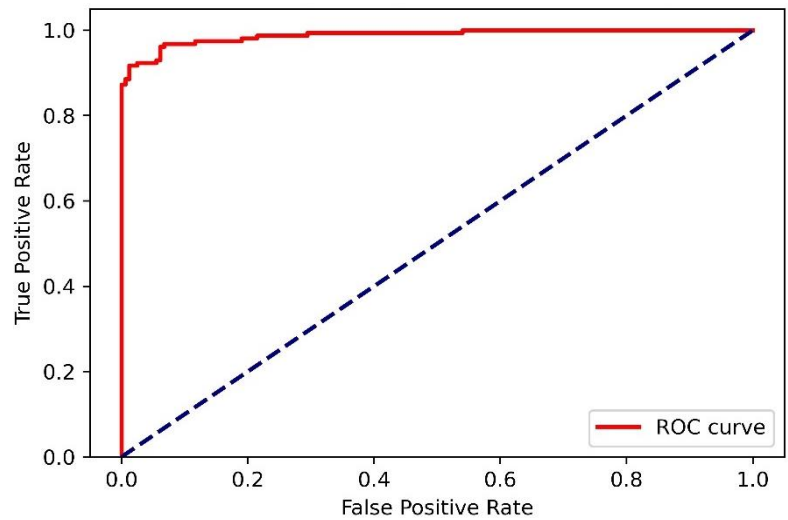

**Figure 6.** ROC plot for the HND model.

|  | Actual | |
|---|---|---|
|  | node | non-node |
| **Predicted** node | 314 | 0 |
| **Predicted** non-node | 7 | 138 |

**Figure 7.** Confusion matrix of the HND model.

## 4. DISCUSSION

The aim of this work was to detect nodules with a low false-negative rate. To this end, the YOLO model was re-trained to have very high sensitivity for nodule detection (using a low confidence level) with high false positive rates. To reduce the rate of false positives, the 3D CNN network was developed on the outcomes of the YOLO model. Through using the 3D CNN model, the accuracy of object detection models increased dramatically, making them a highly promising tool for predicting lung cancers. The HND model (YOLO 0.3 + 3D CNN) was able to accurately detect the nodules in the lung with accuracy and precision of 98.4% and 100% respectively.

When using the confidence level of 0.5 for the YOLO model, 37% of nodules were missed at the first stage. Thus, the confidence level was reduced to 0.3 to increase the sensitivity of the model to detect the entire suspicious nodules (though the model resulted in high false positive rates). By reducing the confidence level of the YOLO model to 0.3, the entire nodules were detected (100%).

Nguyen et al. [28] used a Fast R-CNN (Region-Based Convolutional Neural Networks) as well as a 2D network in order to reduce the number of false positives. As a result, they achieved an accuracy of 95.7% and a sensitivity of 93.8%. Moreover, in a study conducted by Agnes et al. [32], an UNet-based model combined with the pyramid dilated convolutional LSTM resulted in a sensitivity of 93%. In another study [30], Fan et al. used an R-CNN model and 3D Resnet (consisting of 50 deep layers) for nodule detection, and they achieved a sensitivity of 93.6%, while the HND model in this study exhibited a sensitivity (recall) of 97.8%. The 3D CNN classifier led to superior results while having a lower number of layers (17 vs 50), which could be justified by the better convergence and less likelihood of overfitting. George et al. [39] developed an end-to-end process through analyzing the entire input image for lesion detection and classification. They achieved a precision of 89%, which is better than the YOLO model in this work but inferior to the overall precision of the HND model. The increase in precision indicates that the number of false positives has been reduced owing to the promising performance of the 3D CNN classifier. Huang et al. [41] applied a single-stage 3D-YOLOv3 model and achieved a sensitivity of 96.2% which is inferior to what we observed in this study.

This work demonstrated that hierarchical networks would provide an efficient pipeline for nodule detection in the lung. Additionally, the 3D CNN classifier can be used in conjunction with other algorithms, such as segmentation frameworks, to verify the outputs and enhance the overall accuracy of the model.

## 5. CONCLUSION

The objective of this study was to present a hierarchical approach for detecting and classifying pulmonary nodules from CT images. This approach involved two stages; wherein in the first stage, the entire CT image was fed to the YOLO model. The YOLO detected almost the entire suspicious nodules with high sensitivity. Afterward, in the second stage, the 3D CNN classifier found the false positive cases (non-nodules) and dramatically improved the overall accuracy of the model. An accuracy of 98.4% and AUC of 98.9% were achieved.